\documentclass[apjl]{emulateapj}
\usepackage{times}
\usepackage{natbib}
\shorttitle{The horizontal internetwork magnetic field}
\shortauthors{Steiner et al.}

\begin{document}

\title{The horizontal internetwork magnetic field: numerical 
simulations in comparison to observations with Hinode}

\author{O. Steiner and R. Rezaei}
\affil{Kiepenheuer-Institut f\"ur Sonnenphysik, Sch\"oneckstrasse 6,
        D-79104 Freiburg, Germany}
\email{[steiner,rrezaei]@kis.uni-freiburg.de}
\author{W. Schaffenberger}
\affil{Physics and Astronomy Department, Michigan State University, 
        East Lansing, MI 48824}
\email{schaffen@pa.msu.edu}

\altaffiltext{1}{Marie Curie Intra-European Fellow of the European Commission}

\and

\author{S. Wedemeyer-B\"ohm\altaffilmark{1}}
\affil{Institute of Theoretical Astrophysics, P.O. Box 1029, Blindern,
        N-0316 Oslo, Norway}
\email{sven.wedemeyer@astro.uio.no}

\begin{abstract}
Observations with the Hinode space observatory led to the 
discovery of predominantly horizontal magnetic fields in the photosphere 
of the quiet internetwork region. Here we investigate realistic numerical 
simulations of the surface layers of the Sun with respect to 
horizontal magnetic fields and compute the corresponding polarimetric 
response in the \ion{Fe}{1}~630~nm line pair. We find a local 
maximum in the mean strength of the horizontal field component 
at a height of around 500~km in the photosphere, where it
surpasses the vertical component by a factor of 2.0 or 5.6, depending 
on the initial and boundary conditions. From the synthesized Stokes 
profiles we derive a mean horizontal field component that is, respectively, 
1.6 and 4.3 times stronger than the vertical component. This is a 
consequence of both the intrinsically stronger flux density of, and the 
larger area occupied by the horizontal fields. We find that convective 
overshooting expels horizontal fields to the upper photosphere, making 
the Poynting flux positive in the photosphere, while this quantity is 
negative in the convectively unstable layer below it.
\end{abstract}

\keywords{Sun: photosphere --- Sun: magnetic fields --- MHD --- polarization --- turbulence}

%%%%%%%%%%%%%%%%%%%%%%%%%%%%%%%%%%%%%%%%%%%%%%%%%%%%%%%%%%%%%%%%%%%%%%%%%%%%%%%%%%%

\section{Introduction}
\label{sect1}

Recent observations with the spectropolarimeter of the 
Solar Optical Telescope (SOT) onboard the Hinode space 
observatory \citep{kosugi+al07} indicate that the quiet internetwork 
region (the inner regions of supergranular cells of the quiet Sun) 
harbors a photospheric magnetic field whose  mean flux density of the
horizontal component considerably surpasses that  of the vertical 
component \citep{lites+al07,orozco+al07,lites+al08}. According to 
these papers, the vertical fields are concentrated in the intergranular 
lanes, whereas the stronger, horizontal fields  occur most commonly at 
the edges of the bright granules, aside from the vertical fields. 
In a gravitationally stratified atmosphere,  vertical magnetic flux 
concentrations naturally develop a horizontal component as they expand 
with height in a funnel-like manner. Indeed, \cite{rezaei+al07} found
funnel shaped magnetic elements in the internetwork from the same
Hinode data. However, the newly discovered 
horizontal fields also occur apart from 
vertical flux concentrations and seem to cover a larger
surface fraction than the vertical fields. 

Regarding numerical simulations, \citet{ugd+al98}, note: 
``we find in all simulations also strong horizontal fields above 
convective upflows'', and \citet{schaffenberger+al05,schaffenberger+al06} 
find frequent horizontal fields in their three-dimensional simulations, which 
they describe as ``small-scale canopies''. 
Also the 3-D simulations of \citet{abbett07} display 
``horizontally directed ribbons of magnetic flux that permeate the model 
chromosphere'', not unlike the figures shown by \citet{schaffenberger+al06}.
More recently, \citet{schuessler+voegler08} find in a three-dimensional 
surface-dynamo simulation ``a clear dominance of the horizontal field in 
the height range where the spectral lines used for the Hinode observations 
are formed''.  

Here we report on the analysis of existing and new
three-dimensional magnetohydrodynamic computer
simulations of the internetwork magnetic field aiming at
following questions: Does a realistic simulation of
the surface layers of the Sun intrinsically produce horizontal
magnetic fields and can their mean flux density indeed surpass
the mean flux density of the vertical field component? What is
the polarimetric signal of this field and how does it compare
to measurements with Hinode?

In the following we explain in Sect.~2 the details of two simulations
and present results
that answer the first two of the above questions in Sect.~3. In Sect.~4, 
we synthesize Stokes profiles 
and compare them to measurements from Hinode. Conclusions 
follow in Sect.~5.

%%%%%%%%%%%%%%%%%%%%%%%%%%%%%%%%%%%%%%%%%%%%%%%%%%%%%%%%%%%%%%%%%%%%%%%%%%%%%%%%%%%

\section{Two simulation runs}
\label{sect2}

We have carried out two runs, run v10 and run h20, which significantly 
differ in their initial and boundary conditions for the magnetic field. 
This enables us to judge the robustness of our results with respect to 
magnetic boundary conditions. Both runs are carried out within a common 
three-dimensional computational domain extending from 
1400 km below the mean surface of optical depth $\tau_{\rm c } =1$ to 
1400~km above it. With this choice we ensure that the top boundary
is located sufficiently high for not to unduly tamper the atmospheric
layers that are in the focus of the present investigation, in particular
the formation layers of the spectral lines used in polarimetric 
measurements with Hinode.  The horizontal dimensions 
are $4\,800\;\mbox{km}\times 4\,800\;\mbox{km}$, corresponding to
$6.6\arcsec\times 6.6\arcsec$ on the solar disk. 
With $120^3$ grid cells, the spatial 
resolution in the horizontal direction is 40~km, while in the vertical 
direction it is 20~km throughout the photosphere and chromosphere.  
Both runs have periodic lateral boundary conditions, 
whereas the bottom boundary is open in the sense that the fluid can 
freely flow in and out of the computational domain subject
to vanishing total mass flux. The upper boundary is ``closed'', i.e., 
a reflective boundary is applied to the velocity.

\begin{figure}
\centering
\includegraphics[width=0.33\textwidth]{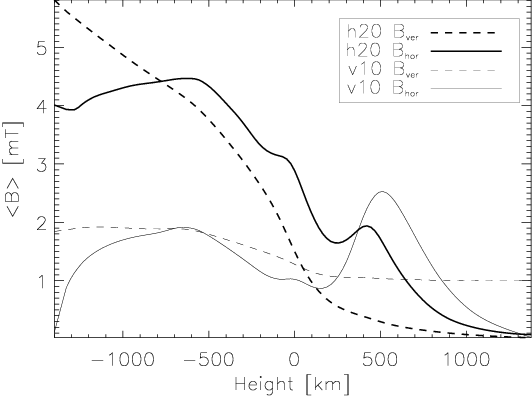}
\caption{Horizontal (solid) and vertical (dashed) absolute field 
strength as functions of height for run h20 (heavy curves)
and for run v10 (light curves).
\label{fig1}}
\end{figure}

Run v10 starts with a homogeneous, 
vertical, unipolar magnetic field of a strength of 1\,mT superposed 
on a previously computed, relaxed model of  thermal convection. 
After relaxation, fields of mixed polarity occur throughout the photosphere 
with an area imbalance of typically 3:1 for fields stronger than 1\,mT. 
The magnetic field in run v10 is constrained to have vanishing
horizontal components at the top and bottom boundary but lines of force
can freely move in the horizontal direction. Although this condition is quite 
stringent for the magnetic field near the top boundary, it still allows the field 
to freely expand with height through the photospheric layers.
The mean vertical net magnetic flux density remains 1\,mT throughout 
the simulation. These inital and boundary conditions might
actually be more appropriate for the simulation of network 
magnetic fields because of preference for one polarity and
the vertical direction.

Run h20 starts without a magnetic field but upwellings that enter the 
simulation domain across the bottom boundary area carry horizontal
magnetic field of a uniform strength of 2\,mT and of uniform direction
parallel to the $x$-axis with them. Outflowing material carries whatever 
magnetic field it happens to have. These boundary conditions are the 
same as used by \cite{stein+nordlund06}. They are  appropriate when
flux ascends from deeper layers of the convection zone, carried by
convective upflows. Starting from a relaxed model of  thermal convection, 
magnetic field steadily spreads into the convective layer 
of the simulation domain and after 600~s slowly begins to 
expand throughout the photosphere, growing in mean absolute 
strength. Reflective conditions apply to the field at the top boundary,
resulting in $\mathrm{d}B_{x,y}/\mathrm{d}z = 0$, $B_{z} = 0$.

The magnetic energy in the box
steadily increases because convective plasma motion strengthens
the magnetic field. After a time of about 2.45~h an equilibrium 
value in magnetic energy  seems to establishes itself when
the mean absolute vertical field strength near the 
surface of optical depth
$\tau_{\rm c}\! =\! 1$ is approximately 1\,mT.
Events of convective plumes pump magnetic fields in the downward 
direction out of the domain so that the mean Poynting flux at the lower
boundary is negative, pointing out of the box.

Runs v10 and h20 have been carried out with an extended version of
the computer code 
CO$^\mathsf{5}$BOLD\footnote{www.astro.uu.se/\~{}bf/co5bold\_main.html} 
that includes magnetic 
fields. The code solves the coupled system of the equations
of compressible ideal magnetohydrodynamics in an external gravity 
field taking non-local radiative transfer into account. For the present
runs, frequency-independent opacities are used, which are also used
for computing the continuum optical depth $\tau_{\rm c}$.
The multidimensional problem is reduced to a sequence of 1-D sweeps
by dimensional splitting. Each of these 1-D problems is solved with a 
Godunov-type finite-volume scheme using an approximate Riemann solver 
modified for a realistic equation of state and gravity. Details of 
the method can be found in 
\citet{schaffenberger+al05,schaffenberger+al06}.

%%%%%%%%%%%%%%%%%%%%%%%%%%%%%%%%%%%%%%%%%%%%%%%%%%%%%%%%%%%%%%%%%%%%%%%%%%%%%%%%%%%

\section{Structure and development of the horizontal magnetic field}
\label{sect3}

Figure~\ref{fig1} shows the horizontally and temporally averaged absolute
vertical and horizontal magnetic field strength as functions of height
for both runs. In run h20, the mean horizontal field strength,
$\langle\sqrt{B_x^2 + B_y^2}\,\rangle$, is larger than the mean strength 
of the vertical component,  $\langle |B_z|\rangle$, throughout the 
photosphere and the lower chromosphere: in run v10 this is the case in 
the height range between 250~km and 850~km. 
It shows a local maximum close to the classical temperature minimum at
a height of around 500~km, where it is 5.6 times stronger than the mean 
vertical field in case of run h20. The horizontal fields also dominate in 
the upper photosphere of run v10 for which case one might expect the 
initial state and boundary condition to favor the development of vertical 
fields rather than horizontal ones. There, the ratio 
$\langle B_{\mathrm{hor}}\rangle /\langle B_{\mathrm{ver}}\rangle$ 
at the location of maximum $\langle B_{\mathrm{hor}}\rangle$ is 2.5.

For the second half of the h20 time series and
in a horizontal section at a height of mean optical depth $\tau_{\rm c} = 1$,
14.2 \% of the total area is covered by horizontal fields
stronger than 5\,mT, while this fraction is 5.1 \% for the vertical 
fields surpassing 5\,mT. At the height of 200~km in
the photosphere and a threshold of 2\,mT the average area fractions
are 25.8 \% and 6.2 \%, respectively. Thus, fields with a horizontal component
larger than a given limit in strength occupy a significantly larger 
surface area than fields with a vertical component exceeding this limit. 
This is a second reason (after inherent strength) why the measured
mean flux density of the horizontal field component may exceed
that of the vertical one.

\begin{figure*}
\centering
\includegraphics[width=0.63\textwidth]{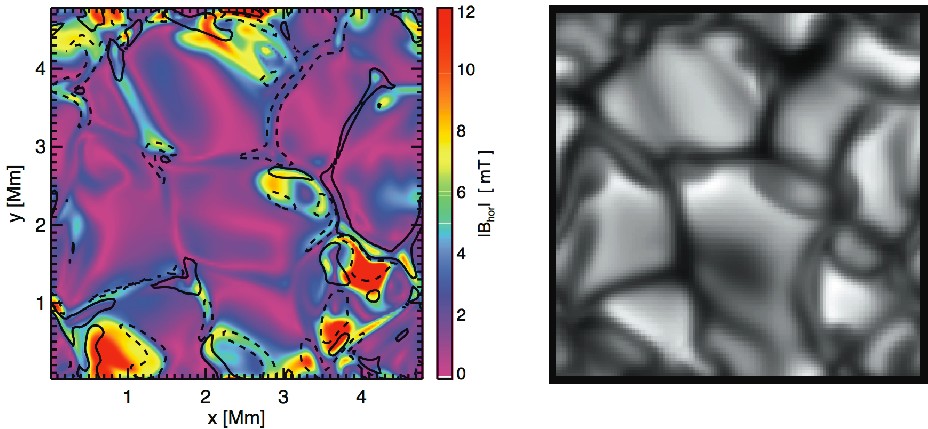}
\caption{Left: Horizontal field strength on the surface of
continuum optical depth $\tau_{\rm c} = 0.3$. The black
curves refer to contours of 2\,mT vertical field strength, where
solid and dashed contours have opposite polarity. Right: Map
of the continuum intensity a 630~nm.
\label{fig3}}
\end{figure*}

Figure~\ref{fig3} (left) shows for a typical time instant in the second half 
of run h20 the horizontal field strength (colors) on the surface of
continuum optical depth $\tau_{\rm c} = 0.3$. Superimposed on the colors
are contours of 2\,mT of the vertical field strength, where solid and dashed 
contours indicate opposite polarity. 
To the right and the lower right of the image center, $(x,y) = (3.1\,,\,2.5)$ and
$ (3.7\,,\,1.6)$, as well as in the middle close to the front side, $ (2.4\,,\,0.3)$,
we can see a frequently occurring event consisting of a ``ring'' of 
horizontal field. It starts to appear as a patch filled with horizontal 
field like to the left front side, $(x,y) = (0.8\,,\,0.3)$, subsequently expanding to 
become a ring. The ring can also be seen in the vertical field 
component, where opposite halves of it have opposite polarity as is visible from 
the indicated  2\,mT contours. This pattern arises from horizontal magnetic 
field that is transported to the surface by vigorous upflows. The field is
anchored in the downdrafts at the edges of a granule or in the 
weaker upwellings of a granule interior. Here, the field is most 
concentrated and hence not only the vertical but also the horizontal 
field is strongest there. As can be seen when
comparing to the continuum intensity image, Figure~\ref{fig3} (right), 
a ring does often not enclose a full granule but only a  part of 
it so that part of the vertical magnetic field occurs within the granule. 
The horizontal field between the crescents of vertical fields of opposite 
polarity covers part of the granule like a cap forming a small-scale canopy
\citep{schaffenberger+al05,schaffenberger+al06}. Events of related
topograpy were observed by \cite{centeno+al07} and \cite{ishikawa+al08}.

As this horizontal field is pushed into the stable layers of the 
upper photosphere by the overshooting convection, it stops rising 
for lack of buoyancy, neither are there vigorous downflows
that would pump it back down again.
Hence, convective flow and its overshooting act 
to expel magnetic flux from the granule interior to its boundaries, 
i.e., not only to the intergranular lanes but also to the upper layers of the 
photosphere.

\begin{figure*}[t]
\centering
\includegraphics[width=0.77\textwidth]{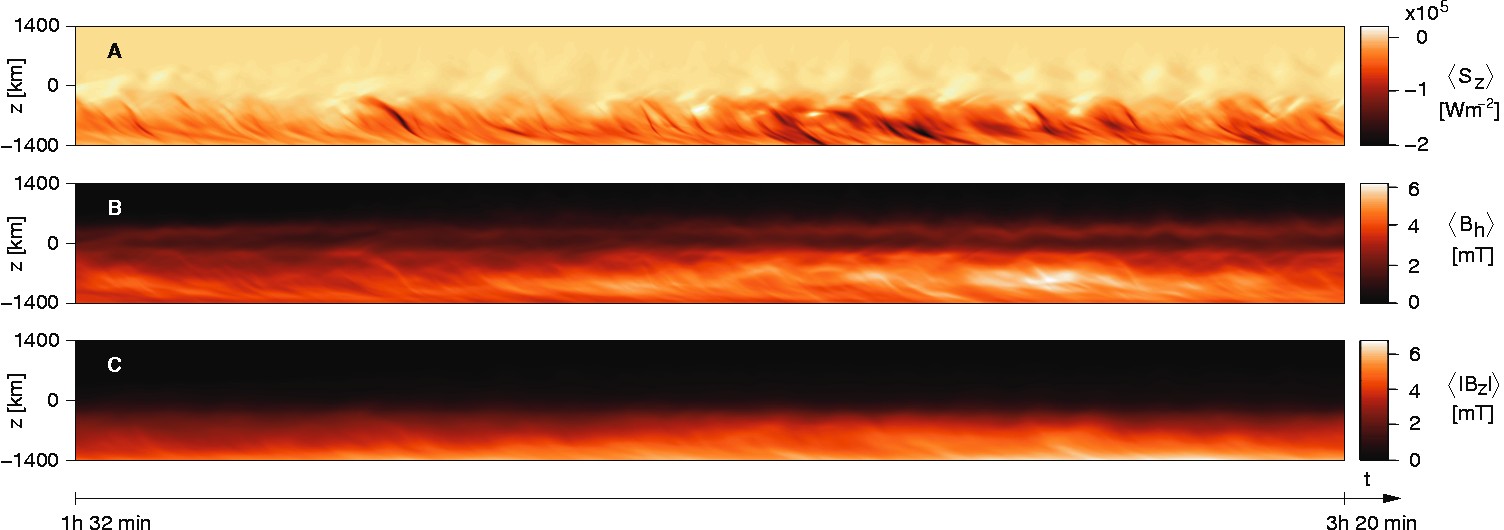}
\caption{Vertically directed Poynting flux, $S_z$, horizontal magnetic
flux density, $\langle B_\mathrm{h}\rangle$, and vertical absolute  magnetic flux
density, $\langle |B_\mathrm{z}|\rangle$ as functions of height and time from run h20.
All quantities are averages in horizontal planes of the three-dimensional
computational box. The temporal average of $\langle S_{z}\rangle$
is maximal $7.4\times 10^2$\,Wm$^{-2}$ (at 200~km) and 
minimal $-5.2\times 10^4$\,Wm$^{-2}$ (at $-800$~km).
\label{fig4}}
\end{figure*}

The surface of $\tau_{\rm c} \approx 1$, which separates the convective
regime from the subadiabatically stratified photosphere, also acts as
a separatrix for the vertically directed Poynting flux, $S_z$, where
\begin{equation}
  \label{eqn_poynting}
  S= \displaystyle\frac{1}{4\pi}\mbox{\boldmath$(B\times 
                      (v\times B))$} .
\end{equation}
This can be seen in Fig.~\ref{fig4} (top), which displays  the horizontally
averaged $S_z$ as a function of height in the atmosphere and time
for the second half of run h20. The conspicuous dark streaks in the 
lower part of the diagram mark events of downflow plumes that carry 
horizontal magnetic field with them, giving rise to $\langle S_z\rangle < 0$.
Differently in the photosphere, where $\langle S_z\rangle$ 
stays mainly positive (bright), due to the transport of horizontal fields in 
the upward direction. They become deposited in and give rise to the  distinct
layer of enhanced horizontal fields in the upper photosphere, clearly visible 
in the middle panel of Fig.~\ref{fig4}. Here again, 
both the transport of horizontal fields downwards in the 
convection zone and upwards in the photosphere are visible.
The bottom panel of Fig.~\ref{fig4} shows the mean vertical field strength that 
monotonically decreases with height at all times.

%%%%%%%%%%%%%%%%%%%%%%%%%%%%%%%%%%%%%%%%%%%%%%%%%%%%%%%%%%%%%%%%%%%%%%%%%%%%%%%%%%%

\section{Comparison with results from the Hinode space observatory}
\label{sect4}

For a reality check we compare the 
Zeeman measurements from the Hinode spectropolarimeter with the
synthesized Stokes profiles of both 630~nm \ion{Fe}{1} spectral lines of 
the two simulation runs. Profiles were computed with the radiative transfer 
code SIR~\citep{sir92, sir_luis} along vertical lines of sight (disk center) with 
a spectral sampling of 2\,pm.  
We then applied a point spread function (PSF) to these `virtual observations':
the theoretical, diffraction-limited PSF of SOT
as well as two other non-ideal PSFs that take additional stray-light 
into account, all evaluated at $\lambda =  630$~nm (see \cite{wedemeyer08}
for details). The following results refer to a PSF obtained by convolution of
the ideal PSF with a Voigt function with $\gamma=5.7\arcsec\!\times 10^{-3}$ 
and $\sigma = 8\arcsec\!\times 10^{-3}$, derived from eclipse data.

For a faithful comparison with the results of \cite{lites+al08}, we subject the 
synthetic profiles to the same procedure for conversion to apparent flux 
density\footnote{Apparent because finite spatial resolution may mask
the true flux density through cancellation of opposite polarization.}
as was done by these authors. Thus, we obtain 
calibration curves for the conversion from the wavelength integrated polarization
signals $V_{\mathrm{tot}}$ and $Q_{\mathrm{tot}}$ to the apparent 
longitudinal and transversal magnetic flux densities 
$|B^{\mathrm{L}}_{\mathrm{app}}|$ and 
$B^{\mathrm{T}}_{\mathrm{app}}$, respectively. Equally, $Q_{\mathrm{tot}}$ 
is the resulting $Q$-profile after transformation to the 
``preferred-frame azimuth'' in which the +$Q$-direction is parallel to the 
projection of the magnetic field vector on the plane of sky, when $U\approx 0$.

Having the calibration curves, we derive spatial and temporal averages 
for the transversal and longitudinal apparent magnetic flux densities,
$B^{\mathrm{T}}_{\mathrm{app}}$ and $|B^{\mathrm{L}}_{\mathrm{app}}|$  
of respectively 21.5~Mx\,cm$^{-2}$ and 5.0~Mx\,cm$^{-2}$ for the second 
half of run h20 and 10.4~Mx\,cm$^{-2}$ and 6.6~Mx\,cm$^{-2}$ for run v10.
Thus, the ratio 
$r=\langle  B^{\mathrm{T}}_{\mathrm{app}} \rangle/
   \langle |B^{\mathrm{L}}_{\mathrm{app}}|\rangle = 4.3$ 
in case of run h20 and $1.6$ in case of run v10. \cite{lites+al08} obtain
from Hinode SP data 
$\langle |B^{\mathrm{T}}_{\mathrm{app}}|\rangle = 55$~Mx\,cm$^{-2}$ and 
$\langle |B^{\mathrm{L}}_{\mathrm{app}}|\rangle = 11$~Mx\,cm$^{-2}$ resulting in 
$r = 5.0$.

While $\langle B_{\mathrm{hor}}\rangle /\langle B_{\mathrm{ver}}\rangle = 5.6$
and $2.5$ for run h20 and v10, respectively, at the location of maximum 
$\langle B_{\mathrm{hor}}\rangle$, the above quoted lower ratios result because
the main contribution to the Stokes signals does not come from this 
height but rather from the low photosphere, where the two components
differ less (see Fig.~\ref{fig1}).
At full spatial resolution, i.e., without application of the PSF, we obtain
from the syntesized Stokes data of run h20 
$B^{\mathrm{T}} = 24.8$~Mx\,cm$^{-2}$ and 
$|B^{\mathrm{L}}| = 8.8$~Mx\,cm$^{-2}$, thus
$r=2.8$.
The higher ratio $r$ when applying the PSF results because of apparent flux
cancellation within a finite resolution element. The vertical component 
is more subject to this effect than the horizontal one because of its smaller 
spatial scale and higher intermittency. This indicates that the predominance 
of the horizontal component decreases with increasing spatial resolution
and that spatial resolution is a fundamental parameter to take into 
account when interpreting measurements of field inclinations 
\citep{orozco+al07}. The probability density for the field
inclination at full spatial resolution of the simulation shows on
the surface $\tau_{\rm c}=0.01$ a flat (isotropic) distribution 
in the range $\pm 50^{\circ}$ from the horizontal direction.

%%%%%%%%%%%%%%%%%%%%%%%%%%%%%%%%%%%%%%%%%%%%%%%%%%%%%%%%%%%%%%%%%%%%%%%%%%%%%%%%%%%

\section{Conclusions}
\label{sect5}

We have carried out  two simulations of magnetoconvection
in the surface layers of the quiet internetwork region of the solar atmosphere.
The simulations greatly differ in their initial state and boundary conditions for
the magnetic field, but otherwise they both equal
faithfully reproduce properties of normal granulation 
(as the magnetic field is weak). The top boundary is placed in the middle
chromosphere (at a height of 1400~km) far away from the photospheric 
layers.

Both simulations intrinsically produce a horizontal magnetic field throughout
the photosphere and lower chromosphere with a mean field strength that 
exceeds the mean strength of the vertical field component at the same height 
by up to a factor of 5.6. The strength of the horizontal field component shows 
a local maximum close to the classical temperature minimum near 500~km 
height (which largely escapes measurements with the 
\ion{Fe}{1} 630~nm line pair). 
Fields with a horizontal component exceeding a certain limit in strength 
occupy a significantly larger 
surface area than fields with a vertical component exceeding this limit.

This horizontal field can be considered a 
consequence of the flux expulsion process \citep{galloway+weiss81}: in the 
same way as magnetic flux is expelled 
from the granular interior to the intergranular lanes, it also gets pushed 
to the middle and upper photosphere by overshooting convection, where
it tends to form a layer of horizontal field of enhanced flux density,
reaching up into the lower chromosphere.
Below the surface of $\tau_{\rm c} \approx 0.1$, 
convective plumes pump horizontal magnetic field in the downward 
direction. Hence, this surface acts as a separatrix 
for the Poynting flux, which is mainly directed upwards above it
and in the downward direction below it.

The response of this field in linear and circular polarization 
of the two neutral iron lines at 630~nm yields a ratio
$\langle  B^{\mathrm{T}}_{\mathrm{app}} \rangle/
 \langle |B^{\mathrm{L}}_{\mathrm{app}}|\rangle = 4.3 $ in case of
run h20 (which, according to Sect.~2, we deem better to represent 
the conditions of 
internetwork regions). This is close to the  measurements of
\cite{lites+al08}, which indicate a factor of 5. Errors may
come from the straylight produced by the spectrograph and
polarization optics that was not taken into account with our PSF,
the difference in mean absolute flux density (run h20, has
only about half the measured value [see Sect.~\ref{sect4}]), 
the frequency-independent treatment of radiative transfer,
lacking spatial resolution, but also natural fluctuations.
The predominance of the horizontal component, may possibly
only exist on a scale comparable to or less than the spatial 
resolution of SOT. At full spatial resolution of the simulation we
obtain a ratio of 2.8 instead of 4.3.

\acknowledgements
The authors thank B.W.~Lites for providing the calibration software,
R.~Hammer and M.~Sch\"ussler for detailed comments on a draft 
version of this paper, and L.R.~Bellot Rubio for helping greatly improving it.
This work was supported by the Deutsche Forschungsgemeinschaft 
(SCHM 1168/8-1).

%%%%%%%%%%%%%%%%%%%%%%%%%%%%%%%%%%%%%%%%%%%%%%%%%%%%%%%%%%%%%%%%%%%%%%%%%%%%%%%%%%%

\clearpage

\end{document}